\documentclass[10pt,journal,compsoc]{IEEEtran}
\usepackage{rotating}
\usepackage{graphicx}
\usepackage[table,xcdraw]{xcolor}
\usepackage{tikz}
\usepackage{capt-of}
\usepackage{algorithm}
\usepackage[utf8]{inputenc}
\usepackage[noend]{algpseudocode}
\usepackage{epsfig,array}
\usepackage{amsmath,amssymb}
\usepackage{amsfonts}
\usepackage{graphicx}
\usepackage{color}
\usepackage{epstopdf}
\usepackage{mathtools}
\usepackage{float}
\usepackage{subfigure}
\usepackage{placeins}
\usepackage{balance}
\usepackage{hyperref}
\usepackage{accents}
\usepackage{adjustbox}
\usepackage[margin=2cm]{geometry}
\usepackage{rotating}
\usepackage{ulem}
\usepackage{dblfloatfix}

\usepackage{xspace}

\usepackage{comment}
\usepackage{textcomp}
\usepackage{xcolor}

\ifCLASSINFOpdf
 
\else

\fi

\begin{document}

\title{Preventing and Controlling Epidemics through Blockchain-Assisted AI-Enabled Networks}

\author{\textbf{Safa~Otoum},~\IEEEmembership{Member,~IEEE,}
        \textbf{Ismaeel Al Ridhawi},~\IEEEmembership{Senior Member,~IEEE,}
        and~\textbf{Hussein T. Mouftah},~\IEEEmembership{Life~Fellow,~IEEE}
\IEEEcompsocitemizethanks{

\IEEEcompsocthanksitem S. Otoum is with Zayed University, UAE. \protect E-mail: safa.otoum@zu.ac.ae

\IEEEcompsocthanksitem I. Al Ridhawi is with Kuwait College of Science and Technology, Kuwait. \protect E-mail: i.alridhawi@kcst.edu.kw

\IEEEcompsocthanksitem H. T. Mouftah is with University of Ottawa, Canada. \protect E-mail: mouftah@uottawa.ca

}}
\clearpage\thispagestyle{empty}
\maketitle

\begin{abstract}
The COVID-19 pandemic, which spread rapidly in late 2019, has revealed that the use of computing and communication technologies provides significant aid in preventing, controlling, and combating infectious diseases.
With the ongoing research in next-generation networking (NGN), the use of secure and reliable communication and networking is of utmost importance when dealing with users' health records and other sensitive information. Through the adaptation of Artificial Intelligence (AI)-enabled NGN, the shape of healthcare systems can be altered to achieve smart and secure healthcare capable of coping with epidemics that may emerge at any given moment. In this article, we envision a cooperative and distributed healthcare framework that relies on state-of-the-art computing, communication, and intelligence capabilities, namely, Federated Learning (FL), mobile edge computing (MEC), and Blockchain, to enable epidemic (or suspicious infectious disease) discovery, remote monitoring, and fast health-authority response. The introduced framework can also enable secure medical data exchange at the edge and between different health entities. Such a technique, coupled with the low latency and high bandwidth functionality of 5G and beyond networks, would enable mass surveillance, monitoring and analysis to occur at the edge. Challenges, issues, and design guidelines are also discussed in this article with highlights on some trending solutions.
\end{abstract}

\begin{IEEEkeywords}
Artificial Intelligence, Machine, Learning, Epidemic diseases, Blockchain, Explainable AI, Plug-and-Play AI.
\end{IEEEkeywords}

\IEEEpeerreviewmaketitle

\section{Introduction}
\IEEEPARstart{T}{he} healthcare sector was hit hard in the years 2020 and 2021 due to the spread of COVID-19 worldwide, causing a global pandemic. Hospital beds were completely full in certain cities, and intensive care units (ICU) were overcrowded. The grim scenario has led governments, industries, institutions, and researchers to rethink and adopt new innovative solutions to prevent, control, and combat such infectious diseases through new healthcare models driven by emerging technological innovations. Although some efforts were introduced that relied heavily on the use of IoT devices for symptom monitoring, such as connected Unmanned Ariel Vehicles (UAVs) and device positioning systems, those solutions are simply temporal and do not provide ongoing sustainability, disease preventiveness, patient data privacy and overall system reliability. Therefore, a novel solution that considers the concepts of distributivity (but also provides secure data sharing), self-learnability (but also provides explainability features), and autonomy needs to be considered to revolutionize healthcare systems and achieve new levels of quality healthcare for citizens globally.

Recent developments in wireless communication, Machine Learning (ML), and decentralized transaction processing techniques like Blockchain, have led to significant changes in all governmental and industrial sectors. The emergence of Beyond 5G (B5G) networks is now leading such sectors to allow for intelligent and on-demand service provisioning. Today, AI is already being adopted in a multitude of healthcare applications, aiding specialists in early-stage disease detection, such as diabetes and cancer. For instance, several medical devices combined with AI have been applied to oversee early-stage heart disease, enabling specialists to better monitor and detect any potential life-threatening cases at earlier and more treatable stages.

\begin{figure*}[ht]
    \centering
    \includegraphics[scale=0.5]{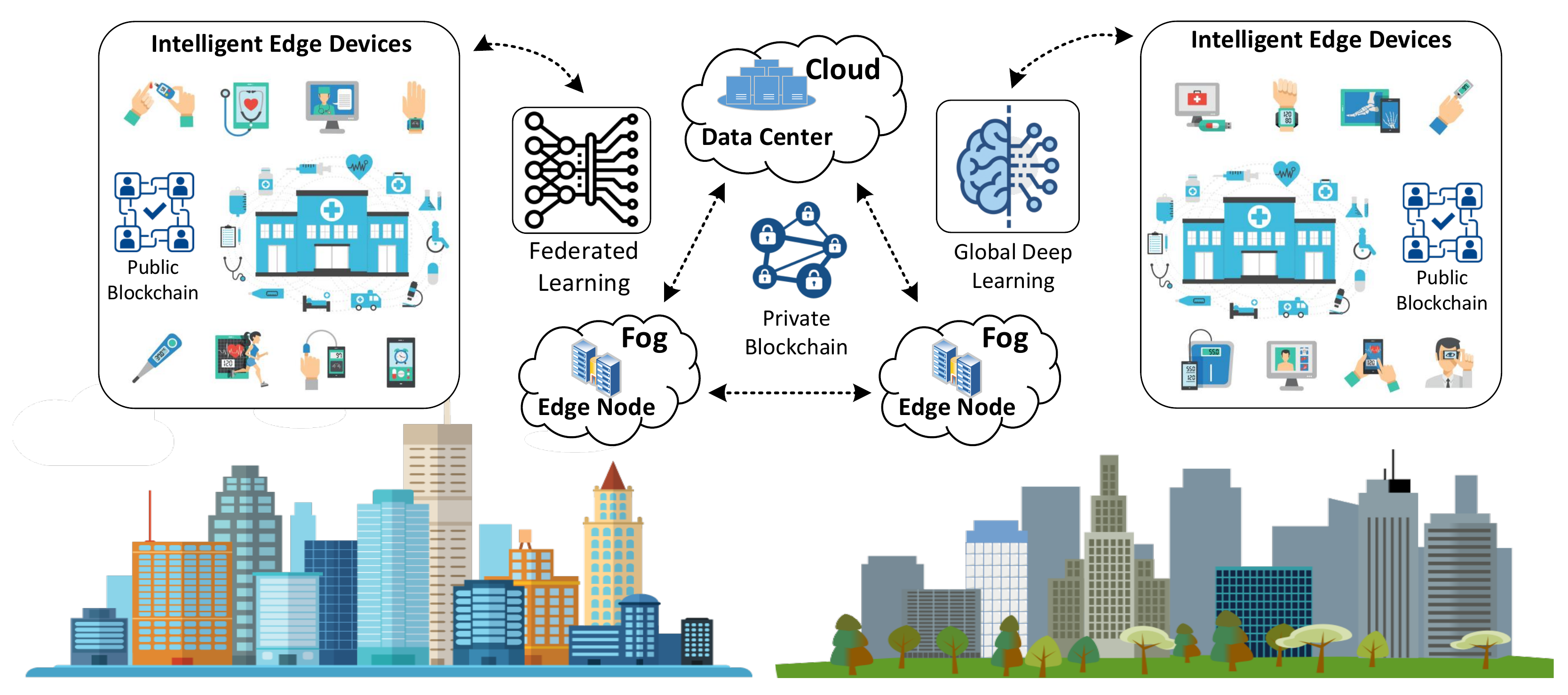}
    \caption{An Overview of a healthcare industry 4.0 scenario that integrates intelligent edge devices, blockchain technology, and machine learning with next generation networking.}
    \label{fig:health4}
\end{figure*}

Although such AI-supported solutions have been adapted in different areas of the health sector, this integration is very limited and would require significant efforts to enable a real shift towards Healthcare Industry 4.0 \cite{health4.0}, illustrated in Figure \ref{fig:health4}. Such shift will enable pre-hospital emergency care in both urban and geographically remote areas. With that said, a huge amount of data will be generated through these applications, allowing for more in-depth training and analysis. But, communicating and sharing this data requires utmost care due to its sensitivity and privacy. Patients need to feel more secure in using applications that would reveal very sensitive and private information and at the same time be able to understand the autonomous diagnosis that may be determined by AI-capable applications. Moreover, the exchange of information across multiple distributed entities will in no doubt lead to better quality level healthcare, but at the same time, such frameworks need to ensure data privacy and integrity.

\begin{table*}[ht]
\centering
\caption{Comparing centralized and federated learning solutions when incorporated in a healthcare framework.}
\label{tab:table2}
\resizebox{\linewidth}{!}{%
{\begin{tabular}{|>{\hspace{0pt}}m{0.119\linewidth}|>{\hspace{0pt}}m{0.355\linewidth}|>{\hspace{0pt}}m{0.466\linewidth}|} 
\hline
 & \textbf{Centralized Learning} & \textbf{Federated Learning} \\ 
\hline
\textbf{Data Ownership} & May remain with the user/source agency.\par{}May be transferred to the centralized entity.\par{}Shared for data control. & Remains at the user/source agency level.\par{}No need for shared data control. \\ 
\hline
\textbf{Privacy/Security} & All data must be communicated with the central entity. & Model updates only.\par{}Data is not communicated with other entities. \\ 
\hline
\textbf{Data Availability} & Frequent data retrieval to ensure accurate model. & Access to data is always available which ensures accurate local model. \\ 
\hline
\textbf{Data Quality} & Collected data from local devices is validated.\par{}More reliable data. & Dependent on the local device.\par{}Some data used in the learning process may be unreliable. \\ 
\hline
\textbf{Performance} & Load is placed on the centralized entity.\par{}Poor performance when considering scalability. & Load is distributed among local devices.\par{}Can accommodate for larger set of user/source agencies. \\
\hline
\end{tabular}
}}
\end{table*}

To ensure reliable early-stage warning, prevention, control and combatant of infectious diseases, rejuvenated health-care systems must consider efficient and reliable storage, processing, and sharing of massive real-time data being collected at the edge. Moreover, when dealing with complex data processing and sharing, such systems must consider and fulfill the diversified security and privacy requirements, which may be enforced either by citizens or government authorities. Given the advances and sophistication of today's edge devices, most of the training and learning can actually be accomplished in a distributed and decentralized manner, namely, Federated Learning (FL) and plug-and-play artificial intelligence (PNP-AI). Such a technique enables the training of a common model to occur with minimal aid of a centralized entity, whilst maintaining all the sensitive data to reside locally (i.e. local health institutions or patient devices). Furthermore, with the aid of Blockchain, secure data sharing and identity security are maintained, and greater transparency for autonomous diagnosis solutions can also be attained using explainable-AI.

\section{Epidemic Disease Management through Intelligent Learning} \label{related}

Adopting ML in healthcare systems enables machines to learn various parameters such as behaviours, symptoms and pathological variables related to patients and diseases. AI-enhanced diagnostics and treatments have been adopted in various healthcare-steered applications. In \cite{10.3390/ijerph15081596}, a framework was developed to optimize the parameters of Deep Learning (DL) techniques in order to predict infectious diseases whilst considering big data. The solution adopted a deep neural network (DNN) along with long-short term memory (LSTM) learning models for predicting three infectious diseases a week ahead of time. In \cite{Kaissis}, a solution for medical imaging applications was developed using next-generation methods for federated, secure and privacy-preserving artificial intelligence. A predictive analysis method for epidemic diseases to identify and locate infected areas by using a back-propagation method was introduced in \cite{doi:10.1063/1.5005397}. The solution-focused on classifying the epidemic disease spreading factors as the elements for weight adjustment on the prediction of epidemic disease occurrences. 

Furthermore, some research works have focused on analyzing the epidemic trends of Severe Acute Respiratory Syndrome (SARS) to validate their solutions. For instance, in \cite{jia2020prediction}, the authors introduced three mathematical models, namely, the Logistic model, Bertalanffy model and Gompertz model. The results have then been utilized to analyze the situation of COVID-19. The authors in \cite{libin2020deep} investigated a deep reinforcement learning mechanism to automatically learn prevention strategies in the context of pandemic influenza. Their work proves that deep reinforcement learning can be an efficient solution for mitigation strategies in complex epidemiological models with a large state space.

\subsection{Federated Learning (FL)}
In FL, a shared ML model is built using distributed data from various devices in which the end-devices train the model using their local data, and subsequently share the model parameters with the coordinating entity, which in essence arranges the model training and gathers the contributions from all devices, without sharing its own local data. A centralized server then builds the global model by aggregating and averaging all collected parameters and broadcasts the new updated model to all end-devices. Each device will then upload its own local model to the server and download the global model. The concept behind decentralized learning is to replace centralized communication with peer-to-peer (P2P) communication. 

FL is very crucial for modernized healthcare systems. The most important advantage of using FL in critical healthcare applications is that patient data is not shared with other devices and nodes, hence, remaining at the patient's device. This will guarantee the desired level of patient data privacy. Furthermore, the use of FL is more convenient and cheaper than traditional centralized learning approaches, due to the minimal need for data updates and the no latency issues given that the model is at the end-users’ devices, allowing for real-time inference. A comparison between the centralized learning approach and that of the federated learning approach when incorporated in a healthcare framework is summarized in Table \ref{tab:table2}.




\subsubsection{Improving FL Efficiency and Effectiveness}
Various tactics can be adapted to improve the effectiveness of FL on the healthcare framework and its efficiency, such as adopting various models to various users, acquiring better optimization algorithms, and upgrading the communication efficiency.

\begin{itemize}
    \item Viral infectious diseases, such as COVID-19, are characterized as viruses that mutate, resulting in different variants and strains. As such, the idea of adapting personalized solutions for different types of end-users becomes very powerful and efficient. The use of different and various models for different end-users would outperform the use of a shared global model. \textit{Personalization through featurization} is a example of this technique, where different end-users run different model gradients (i.e. parameters). Furthermore, \textit{multi-task learning} is another approach that can be accomplished by considering each client's local issue as a distinct task (instead of as a shared one). Any multi-task learning algorithm will be applicable in that case. Additionally, \textit{fine-tuning} is a technique that starts with FL as a single model and applies that model to all end-users. The main difference in fine-tuning is that before the shared model is utilized to make predictions on the end-devices, an additional final training process is considered in order to personalize the model to the users' local data-set.

    \item The goal of FL is to identify a global model that dismisses the risk function over the training data-set, which is, the federation of the data among all end-users. 
    Since the total number of end-devices in FL is significant, the necessity for \textit{client-sampling} algorithms (that necessitate a handful of clients to contribute to each round) is essential. Furthermore, each end-device is expected to participate only once in the model training, which raises the need for state-less algorithms. On the other hand, the optimization algorithms have to be \textit{composable} with other techniques. This is very crucial for a healthcare framework used to combat infectious diseases. Up-to-date data must be gathered to create local models and then shared globally to create accurate models that reflect the situation at different distributed ends. By adapting client-sampling algorithms, faster decisions can be taken to avoid the further spread of disease.

    \item Since wireless edge communication for end-devices may be unreliable and operates at low data rates at certain times, wireless communication could be a crucial bottleneck for FL. This drawback has led to substantial interest in lowering the communication bandwidth, in which a trade-off between communication and accuracy exists in FL. With that being said, with today's 5G wireless communication technology, such a bottleneck will be eliminated to provide enhanced and improved edge communication to support the FL process.
\end{itemize}

Table \ref{tab:DLvsFL} summarizes some of the benefits and drawbacks of adapting FL to preventive healthcare systems over a distributed learning approach.

\begin{table*}[ht]
\centering
\caption{Comparing federated learning and distributed learning in terms of their advantages and disadvantages when adapted to critical infrastructures like healthcare frameworks.}
\label{tab:DLvsFL}
\resizebox{\linewidth}{!}{%
{\begin{tabular}{|>{\hspace{0pt}}m{0.167\linewidth}|>{\hspace{0pt}}m{0.363\linewidth}|>{\hspace{0pt}}m{0.413\linewidth}|} 
\hline
 & \textbf{Distributed Learning} & \textbf{Federated Learning} \\ 
\hline
\textbf{Data Distribution and Accessibility} & Centralized storage of patient data.\par{}Patient data is accessible by end-users. & Patient data is generated and preserved locally.\par{}Decentralized storage of patient data.\par{}Patients do not have access to data stored at other end-devices (either local or shared). \\ 
\hline
\textbf{Data Availability} & All participating patients are assumed to be available. & Only a fraction of patients must be available at any point of time. \\ 
\hline
\textbf{Model Setting} & Training is conducted on a large data-set.\par{}End-users are nodes in a cluster or data-center. & Training is conducted using a local dataset on patients' end-devices. \\ 
\hline
\textbf{Communication} & Communication cost is low.\par{}No wide-area communication. & Communication cost is considered high (wide-area communication).\par{}Consistent connection with end-users. \\ 
\hline
\textbf{Drawbacks} & Computation is processed at the data-center.\par{}Global model is not highly accurate since patient data is not retrieved frequently. & Diversified end-device capabilities may cause issues with computation and communication. \\ 
\hline
\textbf{Advantages} & Few end-user failures. & Local models are constantly accurate, leading to a more accurate global model.\par{}Highly effective for infectious disease scenarios. \\ 
\hline
\end{tabular}
}}
\end{table*}

\subsubsection{Maintaining Patient Data Privacy}
The major issue that faces any healthcare framework is the patient data privacy policy enforced by governments. FL provides a novel technique to address privacy and data governance challenges. It enables ML to occur on non-co-located data. In FL settings, the owner of the local data controls and defines its own governance processes and the associated privacy policies. It also controls entities that may have access to the data and have the ability to revoke accessibility at any moment. With that said, different governmental jurisdictions can enforce their rules and policies on the healthcare framework in a distributed and decentralized manner. End-user FL participants will need to abide by the local governed rules in regards to data sharing.

By moving the model to the data and not vice versa, allows for high-dimensional storage-intense data to remain stored locally without the need for data communication and duplication. If for instance, certain regional government policies dictate that medical data be stored locally within the health region, the local training of the FL process can also occur at the regional level. As such, the proposed healthcare framework is very flexible and can accommodate real-world healthcare scenarios, especially those that require fast and immediate attention, such as in the prevention and control of epidemics.
Some threats may still arise with the use of FL. For instance, an adversary may try to stop a model from being learned or might bias a model under training to generate the adversary's preferred inferences. Therefore, FL requires rigorous protection against such threats. Threat models might arise from the end users' devices or even from the centralized entities. Table \ref{threats_examples} presents some examples of such threats. 

\begin{table*}[ht]
\centering
\caption{FL-based threats that may arise as a result of malicious attacks on the healthcare framework.}
\label{threats_examples}
\resizebox{\linewidth}{!}{%
{\begin{tabular}{|>{\centering\hspace{0pt}}m{0.102\linewidth}|>{\hspace{0pt}}m{0.211\linewidth}|>{\hspace{0pt}}m{0.628\linewidth}|} 
\hline
\multicolumn{1}{|>{\hspace{0pt}}m{0.102\linewidth}|}{\textbf{Targeted Entity}} & \multicolumn{1}{>{\centering\hspace{0pt}}m{0.211\linewidth}|}{\textbf{Data Accessibility}} & \multicolumn{1}{>{\centering\arraybackslash\hspace{0pt}}m{0.628\linewidth}|}{\textbf{Threat Scenario}} \\ 
\hline
\textbf{End-users} & Authorized entities that have access to the end-devices. & Decentralized storage of patient local data may lead to the exploitation of model parameters and training data by attackers.\par{}Malicious devices might have unauthorized access to the learned models.\par{}Some end-users may leverage the benefits of the FL healthcare framework by accessing global models without contributing to the training process.\par{}Unavailability of participant end-users~may yield inefficient results in training the global model. \\ 
\hline
\textbf{Central Server} & Authorized entities that have access to the server. & Attacks on the central server may lead to vulnerability of~initial model parameters, aggregated local models and shared global models.\par{} \\ 
\hline
\textbf{Communication Channel} & Anyone & FL requires~significant amount of communication over the network, where a non-secure communication channel is open to vulnerability through eavesdropping attacks. \\ 
\hline
\textbf{Trained Models} & Developers, data scientists, and software architects. & Malicious end-users may use patient data poisoning attacks on the global model by manipulating the training process.\par{}Malicious end-users may~modify the updated local models before sending them to the central server. \\ 
\hline
\textbf{Global Models} & Anyone & Data reconstruction attacks may lead to global data vulnerability.\par{}Abnormality in local model updates from suspicious end-users may go unnoticed, making the global model vulnerable. \\
\hline
\end{tabular}
}}
\end{table*}

\section{Secure Data Sharing through Blockchain}

A distributed Healthcare framework, where data is collected from patients, filtered, trained, analyzed and processed either locally or through neighbouring nodes, will require a secure and private communication mechanism, that not only is secure but also reliable to ensure data and transaction authenticity and integrity. Additionally, with such a cooperative healthcare framework, in which local and centralized healthcare entities may require real-time access to a collection of patients' records in order to provide timely response to ensure responsive actions towards suspicious infectious disease. Hence, privacy-preserving schemes are essential to ensure proper system integration of all healthcare entities. Data access, the tracking of users' identity, and raw data disclosure are also considered mandatory aspects for any state-of-the-art distributed and intelligent healthcare framework.

Blockchain is highly essential for such cooperative and distributed frameworks, where the healthcare system will be comprised of a significant number of organizations and users having different roles (e.g. patients, healthcare workers, pharmacies, health insurance companies, etc.). It provides a secure mechanism to exchange and store sensitive medical data in a distributed and decentralized manner. Blockchain transactions are considered fast and can provide a mechanism to aggregate policies of different and diversified healthcare entities. This will allow for such diversified entities to cooperate and share a unique model among different healthcare sectors. The adopted blockchain consensus algorithm will make it nearly impossible to modify transaction records, since most of the participants need to verify each block of data \cite{consensus}. This will in essence eliminate cyber-attacks that may occur on data storage sites in a traditional centralized entity scenario.

Many researchers, companies and agencies across the globe started developing blockchain-supported applications to aid in countering the COVID-19 pandemic disease. 
Such applications can validate the continuously changing data, which is quite effective in handling the rapidly escalating COVID-19 situations. In \cite{app1}, the authors launched a Blockchain-supported application, namely Civitas, which is considered a safety system for controlling the impact of COVID-19 by associating people's IDs with Blockchain records in order to verify whether a person in self-quarantining or not. By adopting blockchain to this application, it ensures the security and privacy of patients' data. Another blockchain-supported application, MiPasa, has been introduced in \cite{app2}, which is a data streaming platform that is able to facilitate the sharing of verified health information between people, agencies, and hospitals. It works by gathering the information offered by medical organizations. 

\subsection{Current Trends in Blockchain-Based Healthcare Systems}
Numerous attempts to secure distributed health-care systems using blockchain have been considered lately. Such attempts consider applying the use of blockchain at either the health-care institutions' side (such as hospitals and clinics) or at the patient side. Institution-based blockchain solutions do not involve patients in the secure data sharing and transaction processing aspect, but rather allow them to interact with health institutions to acquire services that have been authorized and authenticated. On the contrary, patient-based blockchain-solutions allow patients to participate in both data sharing and processing transactions. As such, patients are directly involved in the healthcare service provisioning process. Such architectures would require the use of both public and private blockchains. We would assume that involving patients (or even users that contribute and share collected health-related data) requires the use of public blockchains to offer such a decentralized framework. On the contrary, institution-based blockchain solutions would adopt a private blockchain framework solution to allow only authenticated or verified individuals to get involved in data sharing and processing transactions.

\subsubsection{Scalability Issues}
Current blockchain-based healthcare system trends face major issues in regards to the storage of patient health information and learnt models. For instance, in \cite{scalabilityIssues}, the authors proposed a framework for blockchain-based personal health information sharing among healthcare institutions for faster and more reliable disease diagnosis. They integrate two blockchains in their solution, namely, private and consortium blockchains. The private blockchain of healthcare institutions stores raw encrypted patient health information, while the consortium blockchain stores records of the secure indices of patients' health information. Proof of conformance is used for the blockchains as the consensus mechanism. The solution limits the collection and sharing of data among healthcare service providers only and requires patients' approval. Such a technique although is promising, would create storage scalability issues as more data is collected from patients.

\subsubsection{Latency Issues}
Other issues that may face blockchain-based healthcare solutions is the excessive amount of time needed for health authorities to respond to critical infectious disease events through data collected at the patients' side, then shared and communicated to other participants at the edge through public blockchains. By the time this information is shared with the healthcare providers and government authorities, epidemics might have gone out of control. For instance, in \cite{latencyIssues}, the authors tackled the security issues that may arise as a result of transferring and logging data transactions in healthcare systems. Permissioned, consortium-based blockchains are used to execute smart contracts to evaluate information collected from patients. Alerts are triggered for patients and healthcare providers under critical conditions. Although such a solution might provide a mechanism to collect data from patients in real-time, then authenticate this data and communicate it to the health authorities to take action, especially in epidemic related cases, the time needed to filter and analyze the collected information at the healthcare provider side would result in unsatisfactory outcomes.

\subsection{Hybrid Blockchain-Based Healthcare Framework}
\begin{figure}[ht]
    \centering
    \includegraphics[scale=0.29]{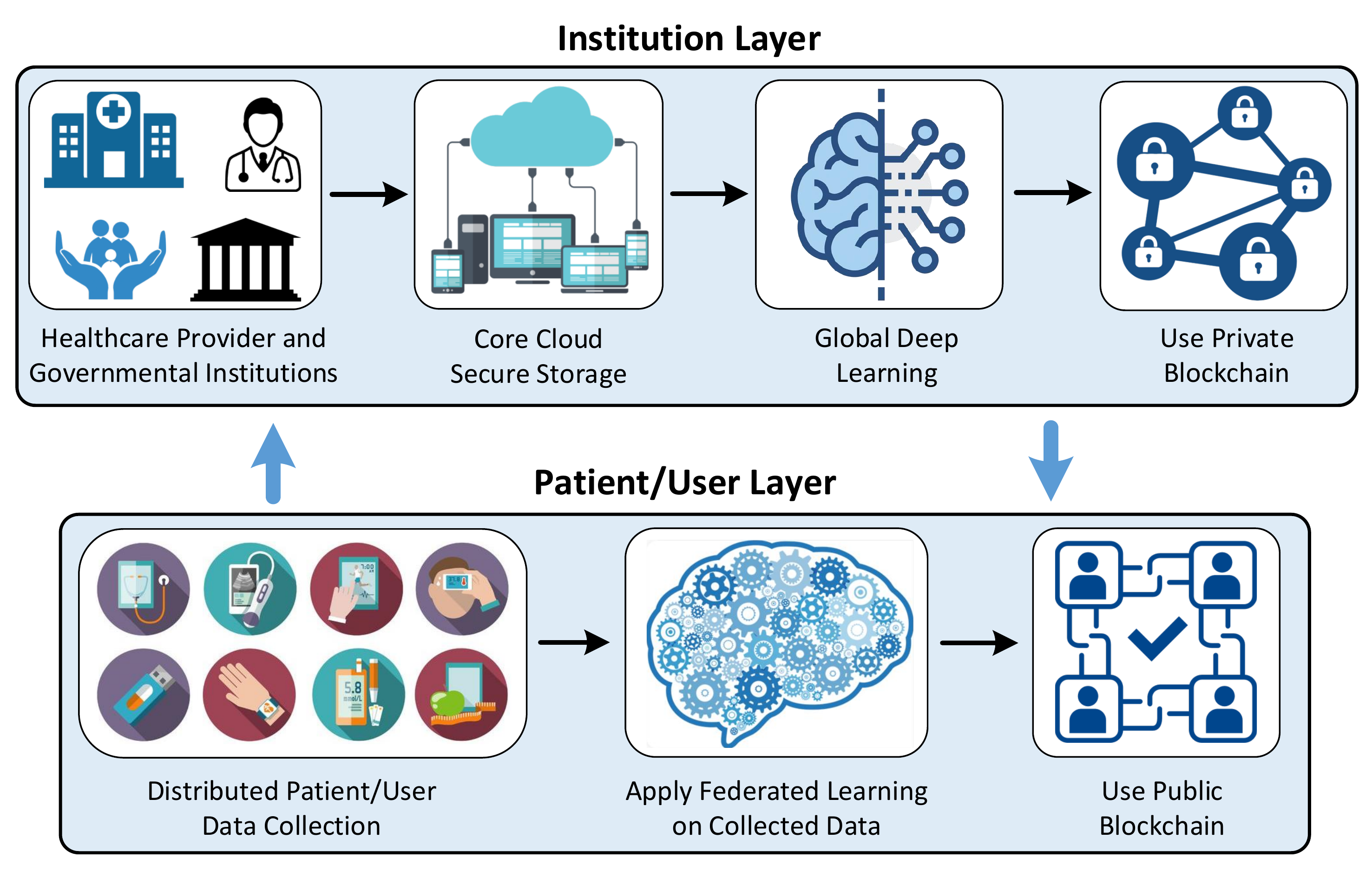}
    \caption{An Overview of a hybrid blockchain solution. The solution adopts a decentralized data collection and federated learning aspect which is applied to a public blockchain for secure sharing and transaction authentication. Healthcare institutions use the collected information and epidemic alerts to provide healthcare services to concerned communities and share data securely with other institutions through private blockchain.}
    \label{fig:layering}
\end{figure}

Hybrid solutions which would assign certain roles to institutions and other roles to users would provide a more optimal strategy towards data collection, sharing, and transaction processing. We envision a healthcare environment that relies heavily on data collection, aggregation, filtering and partial analysis (i.e. Federated Learning) that is achieved using a public blockchain. Data storage and service composition may also be achieved at the edge. 
Moreover, the healthcare framework would also rely on private blockchains to link various medical parties for data sharing to achieve faster and wider service responsiveness, especially for epidemic outbreak scenarios. Figure \ref{fig:layering} depicts a solution that adopts such a hybrid blockchain solution. This approach will not only link the different users at the edge layer, or the institution layer separately, but it will also allow for blockchain system links to occur between patients and doctors. This can be achieved through customized smart contracts, allowing for patients to have on-demand medical data when moving between different healthcare institutions \cite{smartContract}. Adopting FL in the Blockchain-based healthcare system helps in parallelizing the computation capacity of numerous parameters' updates over numerous end-devices. Additionally, this learning environment is an asynchronous learning style where each end-user is independently involved at random occasions which eliminates the need for a global synchronization entity while improving scalability.

There have been a number of recent attempts to address issues related to scalability and latency for blockchain-based healthcare systems. Shamim et al. \cite{shamim} developed a mass surveillance strategy at the network edge to monitor human social distancing, mask-wearing, and body temperature readings. The data is analyzed at the edge by relying on 5G network edge servers. Additionally, data collected in hospitals such as chest x-rays, CT scans, and ultrasound, is also analyzed at the edge. Both data collected from users and hospitals is trained on the cloud using deep learning (DL) models, namely, ResNet50, deep tree, and Inception v3. The trained models are downloaded to the edge during off-peak hours. New data collected either from users or hospitals is sent to the DL models available at the edge. But prior to the data being sent, it is applied on a blockchain to both authenticate and secure the data. Such an approach ensures that the system is scalable in terms of both storage and the number of participants, in addition to providing fast responsiveness for infectious disease diagnosis, results and alerts.

\section{Future Insights}
We envision that in addition to its critical role in patient data exchange among different health entities, blockchain could be used today for setting check-up stations for testing individuals who show symptoms related to an infectious disease. The number of conducted tests in the stations can be managed over the blockchain which can assist in getting reliable reports on the number of conducted tests as well as the number of positive cases. It can help health agencies fight disease spread in areas that have a high number of positive cases. Blockchain can be trusted by healthcare agencies due to its aspect of being immutable. Furthermore, blockchain along with federated learning could be used to store the patients' records, who tested positive to an infectious disease, to contain all their details (e.g. gender, region, age, close contact, etc.) as individual instances, where federated learning could be used to learn and better understand the disease expansion pattern along with facilitating the prediction property. 


\subsection{Explainable and Plug-and-Play AI}
Given the diversity and abundant numbers in both edge and core network devices having advanced processing, storage, and communication capabilities, and given that FL is an optimal choice for decentralized and secure health-care systems capable of adapting to the changing health conditions at different health districts, the adoption of a generalized AI technique that is adaptable to a wide range of devices and applications is highly necessary, namely plug-and-play (PnP) \cite{PnP-AI}. The vision of a simplified and generalized learning solution that may be adapted to any problem and domain, is one that may accelerate the technological advancements in AI and will have a high impact on healthcare provisioning.

Generalized AI refers to a system that can autonomously decide the type of ML algorithm, the training and the processing of the data-set, as well as the reasoning of the data-set selection for optimal feature extraction. Moreover, generalized AI is able to autonomously perform the training and fine-tuning in order to provide an optimal level of generality at the end-devices. Given that all participating end-devices are part of the learning process, PnP-AI would simplify and speed up the FL process. The goal of the PnP-AI solution is to be adaptable to a wide range of edge and IoT devices. PnP-AI is more beneficial when adapted to resource-limited IoT devices, as it gives support to edge devices. The solution is adaptable in the FL context and is highly beneficial, where the learning process is offloaded to end-devices.

In health-related scenarios, the type of diagnosis mechanism needs to first be identified before applying a particular ML algorithm. For instance, in a case where a patient is showing flu-like symptoms, such as high fever, coughing, and low levels of oxygen, which are identified either through body sensors or the use of cooperative devices distributed in a given environment, the type of problem and the input data to be used for training purposes needs to be determined. Such determination of the problem and input data needs to be achieved without user intervention. Then, the training on the data-set, whether performed locally (i.e. distributed) or through FL, the labelling and models are all selected and performed autonomously. Such an approach will require two layers of learning, namely, one that focuses on accurate identification of the configurations and learning mechanisms for the given health-related problem to diagnose the disease, and the other focuses on the problem domain to determine the treatments and actions to be taken, whether locally or by the local healthcare authorities.

With autonomy, comes some disadvantages in terms of system transparency. PnP-AI will further widen the gap between the end-user and the mechanisms involved in decision-making. Both authorities and patients are reluctant to involve the use of AI in medical applications. For AI to be rolled out at a much faster pace in the health-care sector, and due to the sensitivity of the information and decisions/actions made, providing detailed insight into the training, learning, and decision-making process of AI systems is necessary. Explainable AI is a mechanism that would enable such transparency, allowing for AI techniques to be defined, implemented, and interpreted to improve user trust in ML. To achieve a high standard of ethics in AI, explanations of the systems' decisions made by ML algorithms need to be provided to the users. Explainability needs to be adapted at different layers of the system and interpreted for different users according to their expertise. This will ensure user verification, safety and trust of AI results. For instance, decisions made in regards to infectious diseases need to be explained to different categories of healthcare workers and patients.

\begin{figure*}[ht]
    \centering
    \includegraphics[scale=0.5]{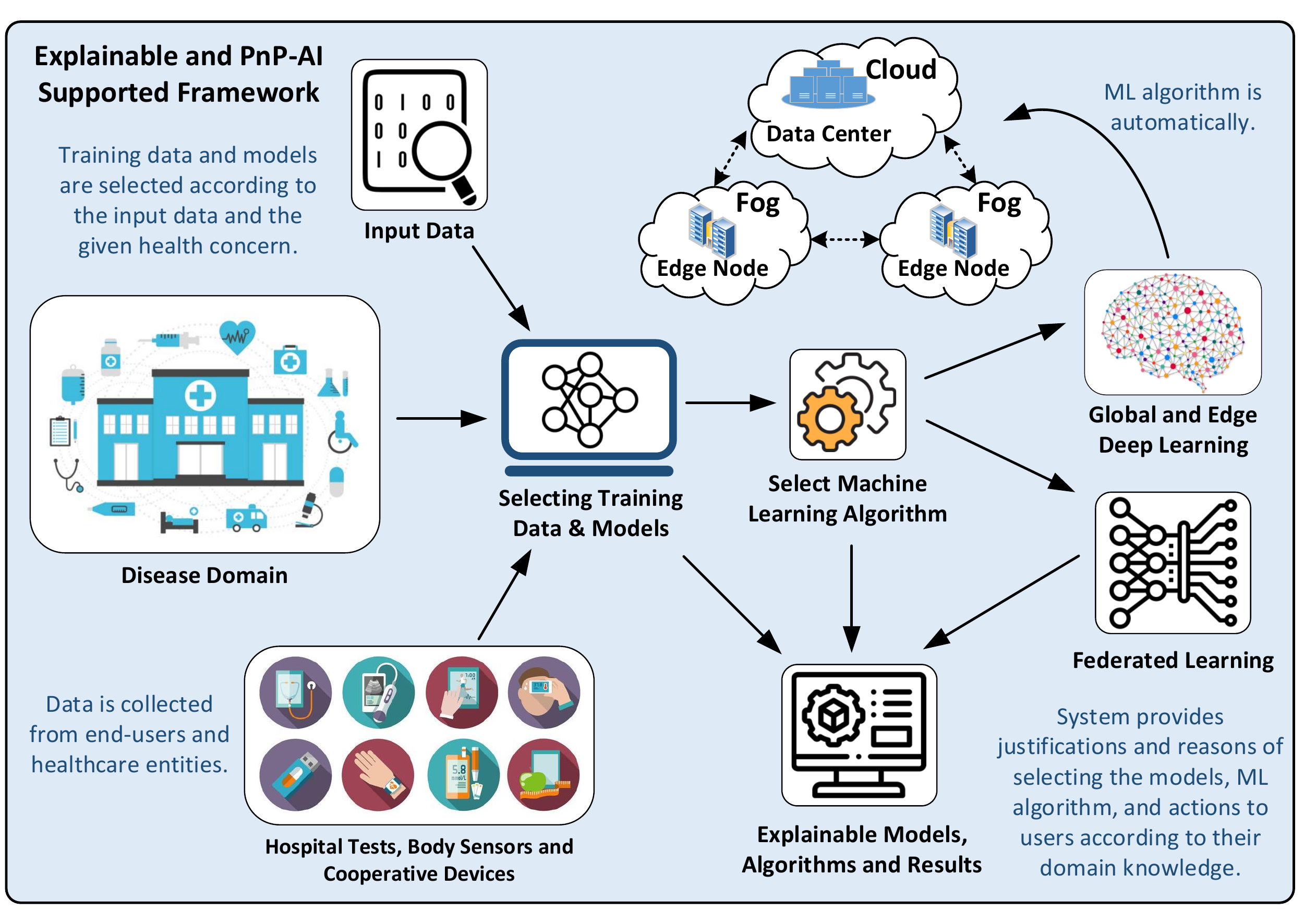}
    \caption{An Overview of an integrated explainable and PnP-AI solution for distributed healthcare frameworks in a cooperative environment to prevent and support fast control and combat of infectious diseases.}
    \label{fig:solutionOverview}
\end{figure*}

Figure \ref{fig:solutionOverview} provides an overview of a healthcare framework that adapts both an explainable and PnP-AI solution. Explainable models need to be created to match the problem domain (e.g. infectious diseases) and the expertise level of the user (e.g. infectious disease specialist, family doctor, nurse, patient, etc.) to describe the data labelling and model selection processes. Since this stage is very critical to ensure accurate diagnosis and actions, the set of users need to be accurately defined. The selection of the ML algorithm will also need to be justified to the user with reasoning to provide more of a glass-box solution style, thus enabling transparency. This will ensure that results presented to the end-user are justified to overcome any safety and health concerns.


\subsection{B5G Network Support}
With the advances in 5G and beyond networks in terms of their mobility and scalability support, mass surveillance and data acquisition at the edge will enable rapid and on-demand health decisions and actions to be considered frequently. Real-time data analysis achieved with the support of roaming end-devices and ultra-fast wireless communication will provide reliable health-decision outcomes in both regional and global avenues \cite{mobility}. Large multi-dimensional datasets and/or locally trained models generated in areas considered highly infectious can be exchanged across 5G networks to generate global models or used by DNNs across different governmental jurisdictions for model training. Medical staff will be able to effectively and promptly respond to urgent cases and have access to regional medical information while roaming across different networks. This will minimize the risk of coming into contact with those that have been infected and will be able to dispatch medical support remotely.

Vertical training, where cooperative machine learning can occur at both the edge and cloud can be conducted more efficiently (in terms of reduced latency and accurate modeling) with the support of B5G. Edge servers with the support of edge caches and edge nodes can provide fast decision-making and monitoring near regional health facilities. Trained models at the edge can be shared with the cloud for further and more accurate global DL. Results can then be rapidly deployed to different jurisdictions globally to ensure synchronized community-based health cooperation. Drug and vaccine manufacturers can determine the frequency and number of vaccines that need to be rolled out to different communities.

\section{Lessons Learned and Concluding Remarks}\label{concluding}
The COVID-19 pandemic has greatly affected and temporally suppressed the majority of healthcare systems globally. This has influenced researchers from different disciplines to develop new techniques and solutions that would enable epidemic discovery ahead of time, remote monitoring, and fast health-authority response. In this article, we envisioned a distributed and decentralized healthcare framework for epidemic diseases' screening, monitoring and rapid reaction. The framework relies on Federated Learning (FL) and Blockchain technology which enable a cooperative and distributed healthcare system. 
We proposed a PnP-AI solution that is capable of adapting to changing device specifications and health conditions. Additionally, we proposed that explainability be adapted to health systems to ensure system decision transparency to all users according to their expertise level. 

For FL-enabled healthcare frameworks to be implemented efficiently, the training process should occur in a distributed manner (i.e. local device training). As such, the adaptation of a PnP-AI solution would ensure proper training and learning to occur at the edge. The implementation of such a solution is indeed complex and is only at its early stages. The concept of achieving a PnP-AI solution would in itself require two layers of learning. One that targets the problem domain and another which targets the management aspect of machine learning. This is considered resource-consuming. As such, the management layer needs to be decentralized and distributed, which would require collaboration between different health entities. Furthermore, the application of blockchain at two levels, namely, user and institution, will provide an opportunity for Electronic Health Records (EHR) to be shared only among those that are given access to the intended data with limited on-chain data storage. The hash of the data is also stored on-chain, thus preserving patient data security. Although such a system is considered complex, we believe that such a proposal will create novel and efficient solutions that are more convenient not only for health authorities but also for patients and citizens.

\section*{Acknowledgment}

This research was supported by the Faculty of Technological Innovation, Zayed University (ZU), under grant number STG-046.

\ifCLASSOPTIONcaptionsoff
  \newpage
\fi

\balance
\bibliographystyle{IEEEbib}
\bibliography{bare_jrnl}

\end{document}